\documentclass[preprint2]{aastex}
\usepackage{graphicx,float,amssymb,natbib}
\def\msun{~M$_\odot$}
\def\kms{~km\,s$^{-1}$}

\shorttitle{Binaries in 30\,Dor}
\shortauthors{Bosch et al.}

\begin{document}
 
\title{Gemini/GMOS search of massive binaries in the ionizing cluster of  30\,Dor. 
}
%\thanks{Based on observations collected at Gemini Observatory}}
%\subtitle{I. Three epoch mid resolution data}
        
% \author[G. Bosch et al.]{Guillermo Bosch$^{1}$ \thanks{IALP (UNLP-Conicet), Argentina},
% Elena Terlevich$^{2}$, Roberto Terlevich$^{2}$ \\
% $^{1}$ Facultad de Ciencias Astron\'omicas y Geof\'isicas, 
% Paseo del Bosque s/n, 1900 La Plata, Argentina  \\
% $^{2}$ INAOE, Tonantzintla, Apdo.\ Postal 51, 72000 Puebla, M\'exico
% }

%\date{in original form  2007 November 23}

%\pagerange{\pageref{firstpage}--\pageref{lastpage}} \pubyear{2008}
%\maketitle

\author{Guillermo Bosch\altaffilmark{1}}
\affil{Facultad de Ciencias Astron\'omicas y Geof\'isicas}
\affil{Paseo del Bosque s/n, 1900 La Plata, Argentina}
\email{guille@fcaglp.unlp.edu.ar}
\and
\author{Elena Terlevich and Roberto Terlevich}
\affil{INAOE, Tonantzintla}
\affil{Apdo.\ Postal 51, 72000 Puebla, M\'exico}
%\label{firstpage}
\altaffiltext{1}{IALP, UNLP-CONICET, Argentina}
\begin{abstract}

If binaries are common among massive stars, it will have important consequences
for the derivation of fundamental  properties  like the cluster  age, IMF and  dynamical mass.

Making use of the multiplexing facilities of Gemini Multi Object Spectrograph
(GMOS) we were able to investigate the presence of binary stars within the
ionising cluster of \object[HD38268]{30 Doradus}.  
From a seven epochs observing campaign at Gemini South we detect a
binary candidate 
rate of about 50\%, which is consistent with an intrinsic 100\%
binary rate among massive stars. We find that single epoch determinations of 
the velocity dispersion
give values around 30 \kms . After correcting the global velocity dispersion 
for the binary orbital motions, the ``true" cluster velocity dispersion is 8.3 
\kms . 
This value implies a virial mass of  about  $4.5 \times
10^5$\msun\ or 8 percent  of the mass calculated using the single epoch value.
The binary corrected virial mass estimate is consistent with photometric mass determinations thus
suggesting that NGC\,2070 is a firm candidate  for a future globular cluster.

\end{abstract}


\section{Introduction}

30~Doradus in the LMC is the nearest available
example of a young and massive starburst cluster. Given its proximity
it is possible to perform a highly detailed study of its stellar component.
The large number of massive stars present in this single cluster allows  
the statistical analysis of several parameters at a level of significance 
that is not available in the local and  smaller Galactic clusters.

Although a large amount of work has been devoted to the study of the 
30-Dor ionizing cluster, some important topics, like the binary fraction among
its massive stars remain relatively unexplored. 
If binaries are common among massive stars, it will have important consequences
for the derivation of fundamental  properties  like the cluster  age, IMF and  dynamical mass.
In particular, the orbital motions of 
massive binary stars in a young stellar cluster can have  
a large impact on the measurement of its global stellar velocity dispersion 
and consequently on the dynamical mass determination.

\cite{1999PhDT.........7B} and collaborators pointed out that the observed radial velocity 
dispersion of stars within the cluster, can be strongly affected by the 
orbital motions of massive binaries. Single epoch mid resolution spectra 
obtained at ESO with the NTT 
\citep{1999A&AS..137...21B} were used to perform a radial velocity analysis
on the stars that conform the ionising cluster of 30\,Dor 
\citep{2001A&A...380..137B}.
The most important results of that analysis can be outlined as follows:

\begin{itemize}
  
\item A high value of the velocity dispersion ($ \sigma \sim 35 $ \kms) was obtained for the OB stars within NGC\,2070, the ionizing 
cluster in 30\,Dor. 
  
\item Estimates of the influence that an underlying binary population could
  have on this determination revealed that orbital motion within the binary
  pair could mimic this large velocity dispersion.

\item From the photometric mass of 30-Dor a rather uncertain value for the virial motions was estimated to be
  around 10\kms.

\end{itemize}

Although these results are consistent with a 100\% binary population,
it is necessary to confirm this assumption with direct
evidence. The issue of binary frequency is still subject to debate, as most of the
statistics on binary stars are biased towards later type stars \citep{2006ApJ...640L..63L}. There is
evidence pointing towards a relatively high number of binaries among early-type stars \citep{1998AJ....115..821M}, {\bf but
previous studies based on single Galactic clusters using similar techniques (such as \citep{2008MNRAS.386..447S}) deal with smaller number of stars, each of them observed on different number of epochs.}

%The presence of violent star formation bursts has been addressed by  \cite{1991IAUS..148..145W} and by 
%\cite{1999a&A...347..532s} and references therein.

%The discovery of massive binaries  could be important in determining the  origin and evolutionary stage of detected  point source X-rays and the %relevance of X-ray binaries and clearly for studies of the Initial Mass Function, topics which are all outside the scope of this paper,where we will %deal with stellar binarity and what it implicates for  the observed kinematics of the stellar cluster.

Here we present a new set of observations of 
NGC\,2070 obtained with the Gemini Multi Object Spectrograph (GMOS) at Gemini
South. These comprise multi object optical spectroscopy of 50 early-type stars
observed at least in six different epochs. The aim is to detect spectroscopic binary 
stars from variations in their radial velocities.

\section{Observations and Data Reduction}
\label{sec:rvobs}

Observations were performed at Gemini South Observatory (Proposals
GS-2005B-Q-2 and GS-2006B-Q-21) using two multislit masks. Targets within each mask were
selected from a previous imaging run with GMOS according to their spectral
types as determined and compiled in \cite{1999A&AS..137...21B}. 
\begin{center}
\begin{table}
 \begin{tabular}{lrr}
\hline\hline
             & MaskI       & MaskII \\  
\hline
07 Sep. 2005 & 2453620.89  & \\
19 Dec. 2005 &             & 2453723.58 \\
20 Dec. 2005 & 2453724.58  & \\
22 Dec. 2005 &             & 2453726.68 \\
23 Dec. 2005 & 2453727.65  & \\
24 Dec. 2005 &             & 2453728.59 \\
29 Nov. 2006 & 2454068.73  & \\
30 Nov. 2006 &             & 2454069.70 \\
24 Dec. 2006 & 2454093.83  & \\
31 Dec. 2006 &             & 2454100.86 \\
01 Jan. 2007 & 2454101.79  & 2454101.73 \\
09 Jan. 2007 &             & 2454109.66 \\
\hline
 \end{tabular} 
\caption{Log of observations. Observing dates are listed in column 1 and Heliocentric Julian dates for masks I and II are shown in columns 2 and 3 respectively.}
\label{tab:log}
\end{table} 
\end{center}
Spectra were obtained during nights of September and December 2005 for the first run and December 2006 and January 2007 
for the second run. Approximate HJD for the 
listed observations are shown in table \ref{tab:log}.
The instrument was set up with the B1200 grating R $\sim 3700$ centred at about 4500 \AA\
which yields a resolution of 0.25\AA\ per pixel at the CCD. Although the wavelength
range varies slightly in MOS spectroscopy, the region from 3900 \AA\ to 5500 \AA\ is covered by our 
spectra. 
The total integration time was split in three to allow for the wavelength
dithering pattern needed to cover the gaps between GMOS CCDs.  Overall 
signal to
noise ratios are above 150.  These ratios are measured on reduced spectra, 
dividing the average value of the stellar continuum by the scatter over the same spectral range.  
 No flux calibration stars were observed, as they were not needed for our purposes.

Data were reduced following standard procedures, using the {\tt GMOS}
reduction tasks within the Gemini IRAF\footnote{IRAF is distributed by the National Optical Astronomy Observatory,
which is operated by the Association of Universities for Research in
Astronomy, Inc., under cooperative agreement with the National Science
Foundation.} package.

\section{Radial Velocities}
\label{sec:rvcal}

\subsection{Zero-point errors}
\label{subsec:zeropoint}

We were able to check for the presence of zero-point errors in our radial
velocity determinations using the nebular lines present throughout the
region. Nebular spectra were wavelength calibrated together with the
stellar spectra, as to provide a strong template to check for variations of our
radial velocities zero-point at different epochs. Nebular emission lines 
have strong narrow profiles, enabling us to cross-correlate 
nebular spectra obtained on different nights, using {\tt fxcor} within IRAF, 
that yields accurate
determinations of radial velocity differences, if present. 
Radial velocities derived for each night ($V_{neb}$) are very
stable when observations on different epochs are compared and their
variations $|\Delta V_{neb}|$ were found to be negligible
( $|\Delta V_{neb}| \leqslant 1.0$\kms, $\sigma_{neb}=3.1$
\kms) which suggests that there is no systematic shift 
introducing spurious variations of stellar radial velocities between epochs.

\subsection{Measurements}

Radial velocities were derived measuring absorption line profiles with the aid
of the {\tt ngaussfit} task within the STSDAS/IRAF package, following a similar
procedure as the one described in \cite{2001A&A...380..137B}. 
This allowed us to derive individual radial velocities 
for each spectral line, and handle
each element (or each ion in the case of He) separately. Stellar radial
velocities were derived using the best set of lines available, according to
the star's spectral type. Although this means we are not using the same set of
lines for the whole sample, we strictly kept the same set of lines for the
same star on different epochs when looking for radial velocity variations.
Uncertainties introduced when fitting individual Gaussians to the absorption
profiles are of the order of 5\kms. 

Table \ref{tab:radvel} lists the complete set of radial velocities determined
for the sample stars. Stars are labelled following the nomenclature of
\cite{1993AJ....106..560P} and data columns include average radial velocity 
(and its 
uncertainty) for each epoch. The errors listed in columns 3, 5, 7, 9, 11, 13
and 15  in Table \ref{tab:radvel}
correspond to the internal error $\sigma_I$ which is the quadratic sum of the standard deviation  of the estimated average from the set of available lines and the minimum uncertainty (3.1 \kms) as derived from high S/N narrow nebular emission lines. Column 16 shows
the ratio between the standard deviation of stellar radial velocity between different epochs and the 
average of $\sigma_I$. Spectral types are given for reference in column 2.

\begin{center}
  \begin{deluxetable}{rlrrrrrrrrrrrrrrr}
\rotate
\tablecolumns{16}
%    \begin{tabular}{rrrrrrrrrrrrrrrr}
%       \hline\hline
%       Id & $V_r$ & $\sigma_{\mathrm{I}}$ & $V_r$ & $\sigma_{\mathrm{I}}$ & $V_r$ &
%       $V_r$ & $\sigma_{\mathrm{I}}$ & $V_r$ & $\sigma_{\mathrm{I}}$ & $V_r$ & $\sigma_{\mathrm{I}}$ & $V_r$ & $\sigma_{\mathrm{I}}$ &
%       $\sigma_{\mathrm{I}}$ & $\frac{\sigma_{\mathrm{E}}}{\langle \sigma_{\mathrm{I}} \rangle}$\\
%       \hline
\tabletypesize{\scriptsize}
\tablewidth{0pt}
\tablecaption{ Stellar
    identifications by \cite{1993AJ....106..560P} are  in column 1 and spectral types from \cite{1999A&AS..137...21B} are included in column 2. 
Columns 3 through 16 include radial velocities and uncertainties (all in \kms) for each
    epoch of observation. Column 17 lists the ratio between epoch-to-epoch variations and the average uncertainties.}
\tablehead{\colhead{Id} & \colhead{Sp. T} & \colhead{$V_r$} & \colhead{$\sigma_{\mathrm{I}}$ } & \colhead{$V_r$} & \colhead{$\sigma_{\mathrm{I}}$ } &
\colhead{$V_r$} & \colhead{$\sigma_{\mathrm{I}}$ } & \colhead{$V_r$} & \colhead{$\sigma_{\mathrm{I}}$ } &
\colhead{$V_r$} & \colhead{$\sigma_{\mathrm{I}}$ } & \colhead{$V_r$} & \colhead{$\sigma_{\mathrm{I}}$ } &
\colhead{$V_r$} & \colhead{$\sigma_{\mathrm{I}}$ } & \colhead{$\frac{\sigma_{\mathrm{E}}}{\langle \sigma_{\mathrm{I}} \rangle}$}}
\startdata
15\,(I)	& O8.5\,V  &	299.1	&	5.5	&		&		&	272.7	&	9.0	&	207.5	&	4.9	&	289.3	&	5.7	&	255.2	&	5.6	&		&		&	5.9	\\
32\,(II) & O9\,IV  &    283.3	&	6.3	&	271.6	&	4.2	&	261.1	&	4.8	&	271.3	&	4.3	&	278.3	&	4.3	&	275.5	&	4.1	&		&		&	1.6	\\
124\,(I) & O8.5\,V  &    266.0	&	4.2	&	239.3	&	3.7	&	209.7	&	6.0	&	268.7	&	3.3	&	243.5	&	3.5	&	261.0	&	5.3	&		&		&	5.1	\\
171\,(I) & O8\,V  &    272.7	&	3.7	&	286.5	&	5.4	&	256.9	&	5.0	&	258.7	&	4.7	&	276.7	&	6.1	&	269.2	&	7.1	&		&		&	2.1	\\
260\,(II) & B0.5:\,V  &    286.0	&	9.1	&	274.7	&	5.9	&	261.0	&	9.4	&	266.6	&	4.2	&	304.5	&	3.8	&	271.5	&	9.1	&	278.4	&	6.1	&	2.3	\\
305\,(II) & B0-B0.5\,V  &    269.5	&	9.7	&	269.5	&	6.1	&	249.9	&	6.4	&	265.4	&	3.3	&	258.9	&	4.8	&	253.1	&	4.9	&	282.5	&	5.3	&	1.4	\\
316\,(I) & O6.5\,IV  &    286.7	&	5.0	&	366.0	&	6.7	&	302.8	&	3.9	&	214.5	&	4.1	&	349.8	&	8.7	&	302.5	&	5.7	&		&		&	9.4	\\
485\,(II) & O8-9\,V  &    277.3	&	5.3	&	260.1	&	5.5	&	255.7	&	8.1	&	278.7	&	3.5	&	251.2	&	5.5	&	259.8	&	6.2	&	269.1	&	4.0	&	2.0	\\
531\,(II) & O8\,V  &    309.3	&	7.1	&	247.6	&	3.7	&	221.2	&	9.3	&	260.9	&	3.7	&	276.9	&	9.0	&	324.6	&	11.3	&	241.7	&	3.5	&	5.3	\\
541\,(I) & O7.5\,V  &    239.2	&	6.2	&	276.0	&	7.0	&	272.4	&	3.4	&	232.8	&	3.9	&	358.5	&	6.2	&	232.1	&	3.9	&		&		&	9.4	\\
613\,(II) & O8.5\,V  &    246.2	&	4.1	&	275.6	&	4.8	&	332.9	&	9.1	&	257.6	&	3.8	&	264.5	&	4.4	&	245.9	&	7.5	&	236.3	&	5.5	&	5.8	\\
649\,(II) & O8-9\,V  &    282.6	&	9.1	&	286.3	&	6.5	&	280.2	&	6.4	&	276.9	&	3.7	&	281.5	&	6.7	&	285.2	&	5.9	&	280.1	&	4.0	&	0.5	\\
684\,(II) & B0:\,I  &    275.7	&	7.9	&	270.4	&	5.4	&	247.8	&	7.8	&	252.7	&	3.4	&	254.8	&	5.0	&	272.5	&	8.7	&	270.7	&	6.4	&	1.9	\\
713\,(I) & O5\,V  &    238.3	&	4.7	&	316.1	&	6.9	&	298.8	&	4.4	&	320.4	&	5.3	&	297.0	&	7.2	&	312.6	&	10.9	&		&		&	4.6	\\
716\,(II)& O5.5\,III(f)  &    274.0	&	3.6	&	273.8	&	7.4	&	266.1	&	4.6	&	271.4	&	3.8	&	275.3	&	4.4	&	279.8	&	4.1	&	277.5	&	3.8	&	1.0	\\
747\,(I) & O6-8\,V  &    182.4	&	3.8	&	213.9	&	5.6	&	306.5	&	5.0	&	331.0	&	4.0	&	248.3	&	9.2	&	306.0	&	7.5	&		&		&	10.1	\\
809\,(II) & O8-9\,V  &    263.8	&	9.4	&	268.7	&	5.5	&	256.0	&	7.9	&	279.3	&	3.8	&	265.2	&	4.0	&	257.6	&	11.7	&	269.1	&	6.1	&	1.2	\\
871\,(II) & O4\,V((f*))  &    275.2	&	7.5	&	272.8	&	6.3	&	268.7	&	8.4	&	281.3	&	3.6	&	282.2	&	8.3	&	289.1	&	4.2	&	278.2	&	8.2	&	1.2	\\
885\,(II) & O5\,III  &    276.1	&	5.1	&	289.1	&	4.1	&	286.3	&	7.6	&	234.8	&	3.2	&	270.5	&	8.1	&	267.2	&	5.3	&	283.5	&	4.8	&	3.5	\\
905\,(I) & O9-B0\,V  &    268.5	&	6.9	&	267.4	&	8.2	&	271.2	&	11.1	&	257.5	&	4.1	&	261.9	&	5.1	&	265.5	&	5.8	&		&		&	0.7	\\
956\,(I) & B1-2\,V  &    268.4	&	6.6	&	284.5	&	6.4	&	239.1	&	3.7	&	265.7	&	6.5	&	270.3	&	7.1	&	274.3	&	6.8	&		&		&	2.5	\\
975\,(II) & O6-7\,V((f))  &    297.4	&	4.2	&	288.8	&	3.7	&	266.7	&	6.5	&	281.1	&	3.6	&	289.2	&	8.9	&	282.9	&	6.0	&	287.2	&	3.8	&	1.9	\\
1022\,(I) & O5:\,V  &    274.3	&	4.5	&	288.9	&	4.9	&	271.1	&	5.0	&	290.0	&	5.4	&	278.4	&	8.2	&	279.5	&	6.6	&		&		&	1.3	\\
1035\,(II) & O3-6\,V  &    287.8	&	11.2	&	278.1	&	7.6	&	263.5	&	8.4	&	277.7	&	4.2	&	273.6	&	4.9	&	285.1	&	3.4	&	281.3	&	4.7	&	1.3	\\
1063\,(II) & O6-7\,V  &    269.5	&	6.6	&	265.3	&	3.7	&	265.9	&	8.9	&	261.6	&	4.1	&	272.3	&	4.2	&	256.5	&	7.2	&	266.4	&	4.9	&	1.0	\\
1109\,(I) & O9\,V  &    274.2	&	5.7	&	293.9	&	8.6	&	278.6	&	4.5	&	278.5	&	3.6	&	253.0	&	4.4	&	248.5	&	8.1	&		&		&	3.0	\\
1139\,(I) & Bo\,V  &    256.3	&	5.4	&	266.2	&	8.0	&	259.1	&	4.1	&	269.4	&	4.2	&	257.1	&	5.5	&	271.7	&	5.4	&		&		&	1.2	\\
1163\,(I) & O4\,If*  &    263.6	&	5.5	&	273.3	&	7.9	&	259.5	&	4.8	&	265.5	&	8.1	&	266.5	&	11.9	&	260.9	&	13.0	&		&		&	0.6	\\
1218\,(II)& O6:\,V  &    294.1	&	3.4	&	283.6	&	4.1	&	273.0	&	7.1	&	287.4	&	3.8	&	281.4	&	3.6	&	297.7	&	10.3	&	294.3	&	4.5	&	1.7	\\
1222\,(I) & O3-6\,V  &    277.8	&	4.7	&	272.7	&	5.9	&	258.2	&	4.8	&	263.4	&	6.0	&	272.3	&	8.4	&	265.9	&	4.6	&		&		&	1.3	\\
1247\,(I) & B0.5\,IV  &    248.7	&	8.8	&	263.2	&	4.3	&	248.1	&	15.3	&	243.5	&	9.1	&	269.7	&	8.4	&	273.3	&	4.4	&		&		&	1.5	\\
1260\,(I) & O3\,V &    289.0	&	5.0	&	294.8	&	5.1	&	279.3	&	5.0	&	294.5	&	4.2	&	291.4	&	9.4	&	297.3	&	5.9	&		&		&	1.1	\\
1339\,(I) & B0-0.2\,IV &    252.7	&	6.1	&	271.8	&	5.2	&	255.0	&	6.3	&	260.8	&	4.8	&	261.0	&	7.1	&	273.7	&	6.6	&		&		&	1.4	\\
1341\,(I) & O3-4\,III(f*) &    287.9	&	7.6	&	300.3	&	6.0	&	301.2	&	4.4	&	305.3	&	5.2	&	267.2	&	12.6	&	304.5	&	13.9	&		&		&	1.8	\\
1350\,(II) & O6\,III(f*)  &    266.1	&	4.7	&	269.3	&	7.5	&	255.9	&	6.9	&	267.1	&	3.8	&	261.7	&	6.5	&	271.6	&	8.9	&	253.0	&	6.1	&	0.9	\\
1401\,(I) & O8\,V  &    248.2	&	3.1	&	260.7	&	3.6	&	279.7	&	5.2	&	282.6	&	4.5	&	321.2	&	10.9	&	237.5	&	6.7	&		&		&	5.3	\\
1468\,(II) & O9.5\,V &    268.2	&	9.5	&	276.1	&	5.7	&	259.9	&	4.4	&	262.7	&	4.3	&	263.0	&	5.8	&	264.1	&	8.4	&	274.1	&	5.9	&	0.9	\\
1531\,(I) & O6.5\,V((f))  &    276.1	&	4.5	&	304.0	&	5.6	&	290.9	&	3.5	&	295.2	&	4.0	&	291.0	&	8.1	&	299.8	&	5.6	&		&		&	1.9	\\
1553\,(I) & O7\,V  &    267.1	&	4.2	&	241.0	&	6.0	&	230.2	&	7.8	&	287.4	&	3.4	&	309.7	&	6.5	&	316.2	&	6.8	&		&		&	6.1	\\
1584\,(II) & B0-1\,V  &    	&		&	281.7	&	5.9	&	250.7	&	6.9	&	277.2	&	4.0	&	299.0	&	7.2	&	283.1	&	36.3	&	293.8	&	6.1	&	1.5	\\
1604\,(I) & B1\,V &    270.9	&	5.8	&	261.3	&	7.6	&	251.0	&	4.0	&	269.4	&	4.1	&	337.1	&	16.2	&	326.7	&	27.3	&		&		&	3.4	\\
1607\,(II) & O7:\,If &    281.7	&	3.6	&	279.7	&	4.7	&	267.2	&	3.7	&	271.8	&	3.9	&	269.1	&	4.5	&	277.9	&	4.1	&	274.4	&	4.7	&	1.5	\\
1614\,(I) & O5-6\,V((f))  &    274.0	&	3.1	&	293.7	&	4.4	&	276.9	&	5.9	&	286.8	&	3.8	&	280.2	&	8.3	&	291.3	&	5.6	&		&		&	1.5	\\
1619\,(I) & O8\,III(f)  &    267.9	&	3.2	&	321.7	&	5.1	&	301.5	&	3.3	&	292.9	&	4.2	&	305.7	&	10.3	&	291.0	&	5.5	&		&		&	3.4	\\
1729\,(II)& B1\,II-III  &    281.3	&	8.5	&	270.8	&	4.6	&	194.2	&	6.4	&	233.2	&	6.4	&	209.2	&	11.5	&	211.0	&	3.6	&	286.1	&	5.2	&	5.2	\\
1840\,(II)& B1\,V  &    262.4	&	8.2	&	270.8	&	4.3	&	273.5	&	5.8	&	245.1	&	5.5	&	271.4	&	3.2	&	251.3	&	5.3	&	296.9	&	5.9	&	2.2	\\
1969\,(I) & B0.7\,IV  &    273.3	&	5.1	&	276.9	&	6.5	&	285.0	&	4.1	&	299.6	&	4.1	&	312.0	&	5.2	&	313.2	&	6.6	&		&		&	3.3	\\
1988\,(I) & B0.5\,V  &    218.3	&	5.3	&	325.2	&	6.2	&	234.5	&	3.6	&	263.0	&	3.3	&	204.2	&	5.6	&	311.6	&	8.0	&		&		&	9.3	\\
% \hline\hline
%     \end{tabular}
%     \caption[]{Stellar
%     identifications by \cite{1993AJ....106..560P} are  in column 1. Columns 2
%     through 7 include radial velocities and uncertainties (all in \kms) 
%     for each
%     epoch of observation. Finally, column 8 lists the ratio between
%     epoch-to-epoch variations and the average uncertainties.}
%     \label{tab:radvel}
\enddata
\label{tab:radvel}
  \end{deluxetable}
\end{center}
\section{Results}

\subsection{Detection of Binary Candidates}

From the different epochs' radial velocities  we can 
check for the presence of variations with time. To quantify this, we follow
the standard procedure of comparing the dispersion of the average radial
velocity for each star with the average uncertainty in the determination of
each radial velocity. This can be done calculating the ``external'' to
``internal'' velocity dispersion ratio
($\sigma_\mathrm{E}/\sigma_\mathrm{I}$) as defined by \cite{1972AJ.....77..138A}. 
Radial velocity variables can then be
easily flagged out as they show $\sigma_\mathrm{E}/\sigma_\mathrm{I}$ above 3,
which is similar to say that the variation in radial velocity is 3\,$\sigma$
above the expected uncertainties.

Binary motion is not the only source for scatter in the determination
of radial velocities for a star. Stellar winds and atmospheric turbulence -- both present in
early type stars -- also
introduce apparent changes in the derivation of radial velocities at different epochs, but in these cases
the internal errors are large too. A clear example of this is Star \#205 in the ionizing cluster of NGC\,6611
\citep{1999RMxAA..35...85B}.

\begin{figure}
  \begin{center}
    \includegraphics[width=.47\textwidth]{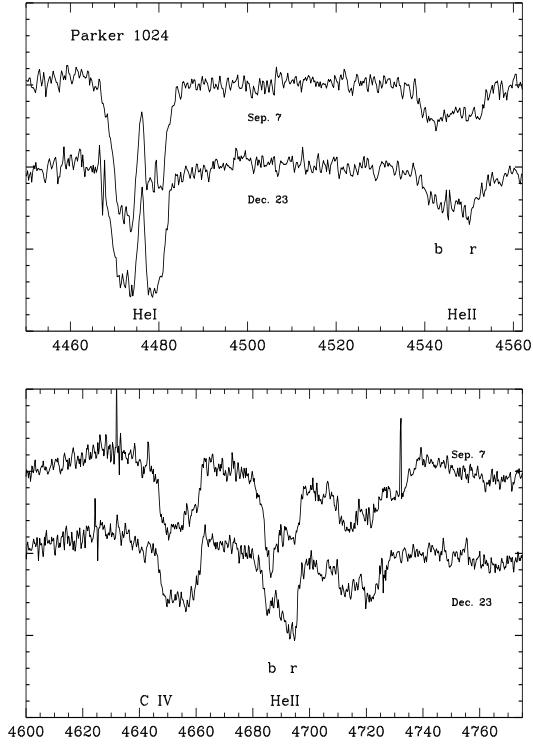}
%no anda    \includegraphics[width=\colwidth]{Parker1024.eps}
    \caption[]{Spectra obtained for Parker 1024 on two different epochs,
    showing the variation of observed profiles for selected absorption
    lines. Both spectra are continuum normalised but shifted in the y-axis by 0.05
    continuum units for
    comparison purposes and identified by their respective observing date 
during 2005.
Relevant absorption profiles are labelled and double troughs
are identified indicating 
their blue (b) and red (r) components. Note the evident changes in the relative intensities
    of the blue and red troughs, particularly for He{\sc{ii}}\,4542\AA\ and He{\sc{ii}}\,4686\AA, suggesting
the binary pair was observed on opposite orbital phases. }
    \label{fig:Parker1024}
  \end{center}
\end{figure}

\begin{figure}
  \begin{center}
    \includegraphics[width=.47\textwidth]{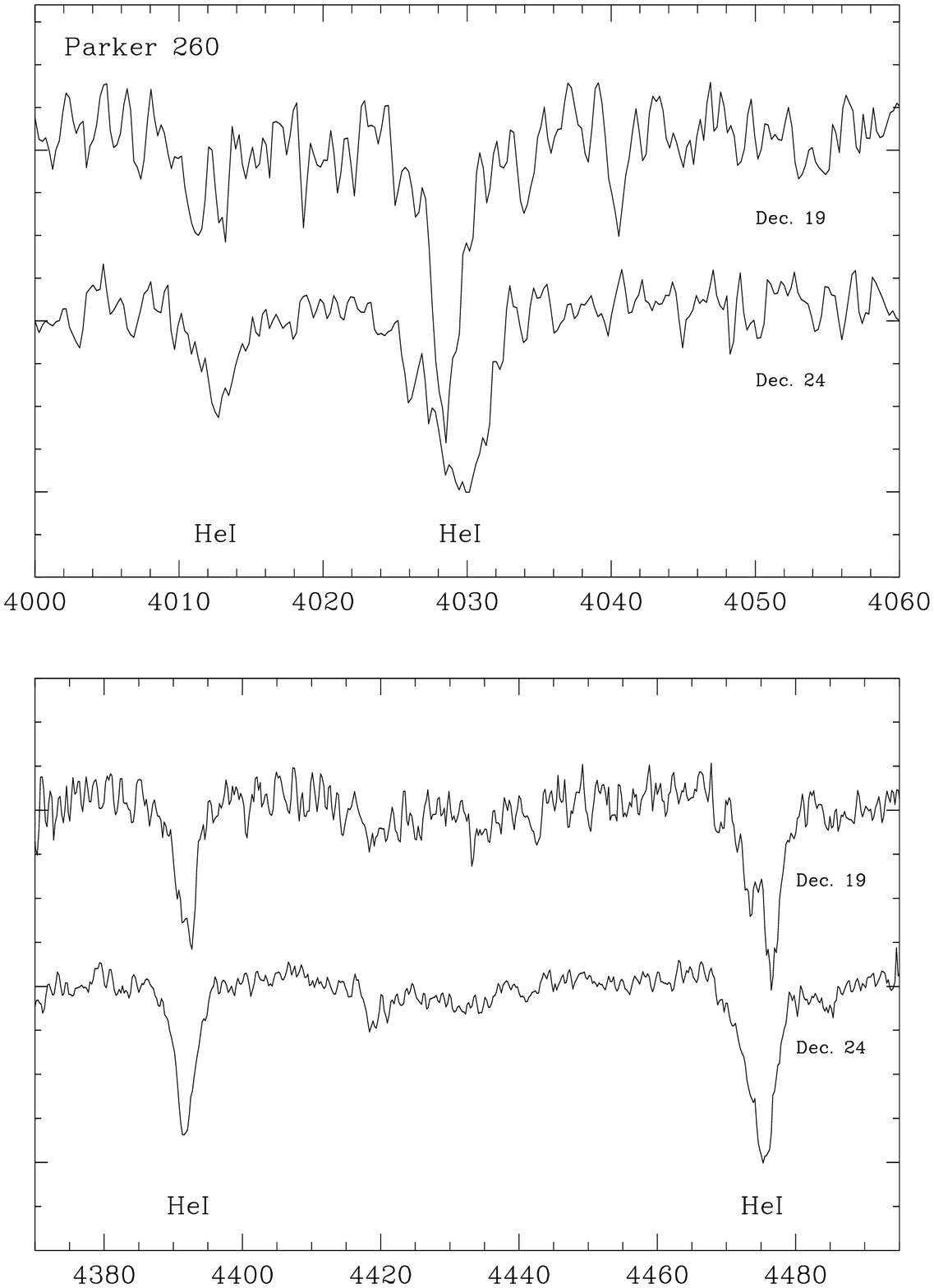}
%no anda    \includegraphics[width=\colwidth]{Parker1024.eps}
    \caption[]{Same as figure \ref{fig:Parker1024} for Parker 260 with a 0.2 continuum units shift between spectra. 
    In this case the profile variations show changes from a well-behaved absorption profile 
    (Dec 24) to an asymmetrical profile (Dec 19). These changes are particularly evident 
    for He{\sc{i}}\,4387\AA\ and
4026\AA\ profiles. Although the most evident changes are seen as a double 
peaked profile in He{\sc{i}}\,4471\AA\ the presence of nebular 
emission contamination in this line cannot be ruled out. }
    \label{fig:Parker260}
  \end{center}
\end{figure}

In addition to the results presented in Table \ref{tab:radvel}, 
the high signal to noise ratio of our spectra has also allowed us to detect 
several double-lined binaries that show evident variation of
their absorption line profiles from epoch to epoch. Examples of these can be
seen in Figures \ref{fig:Parker1024} and \ref{fig:Parker260}.  Individual
radial velocities  are difficult to
measure for these cases, therefore some of these stars are not included in 
Table \ref{tab:radvel}. 
These ten stars (including four present in Table 2) are listed in Table \ref{tab:doublespec}.

An inspection of Table \ref{tab:radvel} shows that 17 out of 46 stars show
radial velocity variations. If we add the stars present in Table 
\ref{tab:doublespec} 
(excluding P613 which already shows radial velocity 
variations in Table \ref{tab:radvel}),
we have our complete set of binary star candidates, which rises to 25 out 
of 52 stars ($\sim 48\%)$.

The mass ratio of binary components, orbital parameters, number of
observations and radial velocity accuracy, can impact the statistics of
detected binaries in star clusters. In an effort to quantify these
effects, \cite{2001RMxAC..11...29B} performed Montecarlo simulations including all the
parameters listed above. Feeding our GMOS data parameters into these
simulations, we find that our detection rate is consistent with a
population of 100\% binaries.

% Interestingly enough, this detection rate is consistent with a
% spectroscopic population of 100\% binaries, when the population characteristics (random orbital plane inclinations, 
% mass ratio distribution favouring $Q=m_1/m_2=1$, period distribution uniform in $\log(P)$ and 
% observational parameters (number of observations and radial velocity accuracy)
% described in \cite{2001RMxAC..11...29B} are set for our GMOS observations. Those
% basic simulations yield an expected 32\% detection rate for the velocity uncertainties
% present in our GMOS data and the number of repeated observations available.
% If
% we were to trust figures to the few percent level, we are detecting even more
% binaries than expected, which might be telling us about the simulated 
% binary population parameters. As found by \cite{2006ApJ...639L..67P}, and with a possible 
% formation scenario devised by \cite{2007ApJ...661.1034K} the distribution of 
% mass
% ratios should be even more concentrated around identical masses for both
% components. 

\begin{table}
  \begin{center}
    \begin{tabular}{ll}
      \hline\hline
      Id & Comment\\
      \hline
      260 & Asymmetrical on 19/12 \\
      613 & He{\sc{ii}} lines look double on 24/12 \\
      674 & Double lines on 22/12 \& 24/12 \\
      805 & Double lined on 23/12 \\
      955 & Looks double on 20/12 \& 23/12 \\
      956 & Looks Asymm. on 23/12 \\
      1024 & Double profile on three epochs \\ 
      1031 & Double lined on 07/09 \& 20/12 \\
      1035 & He{\sc{i}} lines look broad on 24/12 \\
      1938 & Broad lines on 22/12 \\
      \hline\hline
    \end{tabular}
    \caption[]{Stars with evidence of binarity from individual spectra. Stellar
    identifications by \cite{1993AJ....106..560P} are shown in column 1. 
    Column 2 describes the characteristics found among spectral features that
    suggest binary nature.}
    \label{tab:doublespec}
  \end{center}
\end{table}

\subsection{Cluster Kinematics}

After removing the confirmed and candidate binaries, the non-binary population
of our sample decreases to 26 stars, still enough to calculate a representative
value of the stellar radial velocity dispersion (${\sigma_r}$). In calculating it, we must
keep in mind that each radial velocity measurement has its intrinsic error ($\sigma_{\mathrm{int}}$) 
which must be subtracted quadratically from the observed ($\sigma_{\mathrm{obs}}$) velocity dispersion.
The former is the average of the $\langle \sigma_{\mathrm{I}} \rangle$ values over 
the whole sample of non radial velocity variable stars while the latter is directly derived as the 
standard deviation around the average radial 
velocity derived for the same subset of stars in the cluster and it therefore follows that 
${\sigma_r}^2 = {\sigma_{\mathrm{obs}}}^2 - \langle{\sigma_{\mathrm{int}}}\rangle^2$. 
For our sample, $\sigma_{\mathrm{obs}} = 10.3$\kms,
$\langle{\sigma_{\mathrm{int}}}\rangle$ = 6.2\kms, which yield a value of 8.3\kms\ 
for the actual radial velocity dispersion. This seems to confirm the suggestion by 
\cite{2001A&A...380..137B} based on simulations,  that the large values
derived for the stellar velocity dispersion were most probably due to the
presence of binaries. As expected, if we derive the radial velocity dispersion 
from an individual GMOS mask observation for a single epoch,
we find values as high as 30\kms, consistent with what  was previously found 
in Bosch et al. (2001) from single epoch NTT data. 

This is the first time that a radial velocity dispersion is
found for massive stars in NGC\,2070, free from contamination by motion of 
binary stars. We
can therefore safely proceed to compare it with the stellar kinematics expected
for 30\,Dor in order to check if the
cluster is virialised. Following the estimations for the photometric
mass derived in \cite{1999A&A...347..532S}, we find that the stellar mass
within the volume covered by our spectroscopic survey is about $4.5 \times
10^5$\msun. If the stellar velocity distribution is isotropic, we can follow
\cite{1987gady.book.....B} and estimate
the expected radial velocity dispersion as,
$$ {\sigma_r}^2 = \frac{4.5\times10^{-3} \, M}{3 \,R_c} $$ 
where  masses are in \msun, distances in
pc and $\sigma$ in \kms. The velocity dispersion derived for the total mass of
30\,Dor, for a half-mass radius of 14\,pc is about 8 \kms. Our derived value of
8.3 \kms\ is well within the observational -- and sampling -- uncertainties. 
If there are more binaries yet to be discovered
a slight reduction in the observed value is expected 
(although it is unlikely that they would show such large radial
velocity variations that would enhance the current value dramatically).

% \begin{table}
%   \begin{center}
%     \begin{tabular}{lll}
%       \hline\hline
%       Observed ($\sigma_{\mathrm{obs}}$) & Internal
%       ($\langle{\sigma_{\mathrm{int}}}\rangle$) & True ($\sigma_r$)\\
%       \hline
%       10.42\kms & 5.46\kms & 8.87\kms \\
%       \hline\hline
%     \end{tabular}
%     \caption[]{
% Stellar velocity dispersions derived from the binary free sample.
% }
%     \label{tab:sigma}
%   \end{center}
% \end{table}

% \ojo Con una masa creciente con R (casi linealmente) la situaci\'on es bastante
% independiente de la elecci\'on de R, ya que en la expresion de sigma R est\'a 
% en
% el denominador. El valor est\'a definido entonces por el factor de escala en la
% expresi\'on de la masa. De la expresi\'on de Fernando, calcula unas $3 \times
% 10^5$ \msun interiores a 20\,pc. eso nos da unas $4.5 \times 10^5$ \msun 
% para 30\,pc (que es nuestro muestreo). Si le sumamos $1 \times 10^5$ \msun
% calculado para gas ionizado en la regi\'on, con esa masa total nos da que
% esperar\'iamos un $\sigma_r$ de 8 \kms para un radio efectivo (que contiene la
% mitad de la masa) de 14pc. \ojo
% \begin{figure}
%   \begin{center}
%     \includegraphics[width=.47\textwidth]{virial.eps}
% %no anda    \includegraphics[width=\colwidth]{Parker1024.eps}
%     \caption[]{
% Incluyo el dibujito de la evoluci\'on de sigma (y otros) con el radio. S\'olo para
% consumo interno.
% }
%     \label{fig:virial}
%   \end{center}
% \end{figure}

Another interesting issue is that the single stars  population does not confirm
the evidence towards mass segregation that was suggested
in \cite{2001A&A...380..137B}, at least for the
massive stars. However, the fact that massive stars seem to have
a relatively low velocity dispersion cannot rule out dynamical mass segregation
with low mass stars.

\section{Conclusions}
\label{sec:remarks}

 Based on seven epoch observations with GMOS in Gemini South,  
we have presented observational evidence that shows that the
binary fraction among massive stars in NGC\,2070 may be very high. 
Indeed we already detect a 50\% binary candidacy with only
three epoch observations, which suggests that it is only a lower limit.
The evidence pointing towards a high binarity fraction has an important 
effect on the
massive end of the stellar cluster IMF, and shouldn't be overlooked.

If a high binary fraction is common among massive young clusters, virial 
masses obtained from the observed global velocity dispersions in unresolved 
young clusters might be overestimated by a large factor as discussed in
\cite{2008A&A...480..103K}.

Regardless of the origin of the stellar radial velocity variability, the analysis of the
non-variable subset provides an important result regarding kinematics of the
stellar cluster itself.
The radial velocity dispersion 
determined for the 30 Doradus ionising cluster
agrees, within observational errors, with the stellar kinematics expected if
the cluster is virialised and its total mass is derived from the photometric
plus ionised gas masses. This suggests that the stellar cluster is far from
disruption and stands as a firm candidate for a future globular cluster system.

More observations in further epochs should allow us to confirm the binary nature of 
radial velocity variables detected in this work and to derive orbital
parameters and individual masses of members of the binary pairs. The success of GMOS in the
MOS mode makes it a very efficient tool for discovering massive binaries in large
numbers using a relatively small amount of telescope time.

\acknowledgments
We wish
to thank the suggestions and comments given by the referee, Chris Evans, which
improved the final version of this paper. We also acknowledge the invaluable help 
from Rodrigo Carrasco at Gemini South for clearing
our doubts and even providing some scripts that were extremely useful for
endless repetitions of the complete set of reduction procedures from scratch.
We have enjoyed helpful discussions with Daniel Carpintero, Daniel 
Rosa-Gonz\'alez and Guillermo H\"agele.
ET and RT are grateful for a Visiting Professorship and the hospitality of the 
Facultad de Ciencias Astron\'omicas y Geof\'isicas de La Plata 
during a visit when this paper was written.
Based on observations obtained at the Gemini Observatory, which is operated by the
Association of Universities for Research in Astronomy, Inc., under a cooperative agreement
with the NSF on behalf of the Gemini partnership: the National Science Foundation (United
States), the Science and Technology Facilities Council (United Kingdom), the
National Research Council (Canada), CONICYT (Chile), the Australian Research 
Council (Australia), CNPq (Brazil) and SECYT (Argentina)

{\it Facilities:} \facility{Gemini:South (GMOS)}

\bibliographystyle{apj}
\bibliography{30dor_gmos_i}

\label{lastpage}
\end{document}